\documentclass[twocolumn,secnumarabic,amssymb, nobibnotes, aps, prd]{revtex4-1}
\usepackage{amssymb}
\usepackage{amsthm}
\usepackage{amsmath}
\usepackage{bm}

\begin{document}

\title{Exchange symmetry and low-frequency asymptotics of Green's functions in spin s=1 magnets}%
\author{A.V. Glushchenko}%
\affiliation{National Science Center «Kharkov Institute of Physics and Technology», 61108, Kharkov, Ukraine}
\author{M.Y. Kovalevsky}%
\email[M.Y. Kovalevsky: ]{mikov51@mail.ru}
\affiliation{National Science Center «Kharkov Institute of Physics and Technology», 61108, Kharkov, Ukraine}
\author{L.V. Logvinova}%
\affiliation{Belgorod State University, 308015 Belgorod, Russia}
\author{V.T. Matskevych}%
\affiliation{National Science Center «Kharkov Institute of Physics and Technology», 61108, Kharkov, Ukraine}

\date{April 04, 2016}%
\begin{abstract}
The paper contains the description of the dynamics of non-equilibrium processes of spin $s=1$ magnets in an external variable field. We have obtained nonlinear dynamic equations with sources and calculated low-frequency asymptotics of two-time Green's functions for ferro- and quadrupole magnetic states with $SO(3)$ and $SU(3)$ exchange symmetry of the Hamiltonian. It has been shown that for ferro- and quadrupole magnetic states singularities of Green's functions in wave vectors $1/k$, $1/k^2$ and frequencies $1/\omega, 1/\omega^2$ have well-known character. We set the exact form of the magnetic anisotropy of these Green's functions. For states with $SO(3)$ symmetry of the exchange Hamiltonian, we have found Green's functions with quadrupole degrees of freedom and compared them with Green's functions of magnets having exchange $SU(3)$ symmetry.
\end{abstract}
\maketitle
\tableofcontents

\section{Introduction}\label{sec1}
Traditional condensed matter physics deals with rather simple continuous symmetry groups, which include groups of translations or rotations in the space $SO(3)\sim SU(2)$. More complex groups of the unitary $SU(n)$ symmetry with $n>2$ were used in condensed matter physics for description of high-temperature superconductors, low-dimensional semiconductors and high-spin ($s>1/2$) magnets \cite{1,2,3,4,5}. The goal of our research is to elucidate the influence of the exchange interaction symmetry on collective properties of magnets with spin $s=1$ and calculate Green's functions in the hydrodynamic limit, when the wave vector and frequency tend to zero: $k\rightarrow0$, $\omega\rightarrow0$.

Papers \cite{6,7,8,9,10,11,12,Fridm} have studied magnetic equilibrium states with the spin $s=1$ and considered Hamiltonian models with the exchange $SO(3)$ or $SU(3)$ symmetry. In these physical systems unusual states due to magnetic degrees of freedom, which do not change sign under time reversal may appear. These degrees of freedom include the quadrupole matrix, if the exchange interaction has the $SU(3)$ symmetry and nematic order parameter. The Landau-Lifshitz equation \cite{13} describes the evolution of magnets only by means of the magnetization vector. It is well established for magnetic dielectrics with the spin $s=1/2$. Systems with the spin $s=1$ require expansion of the set of magnetic degrees of freedom. In order to describe non-equilibrium processes in pure quantum states of such magnets, just four magnetic degrees of freedom are sufficient \cite{9,14,15}. In papers \cite{16,17,18,19}, more general dynamic equations have been constructed and valid for mixed quantum states. For normal multi-sublattice and degenerate single-sublattice magnets with the spin $s=1$ and the Hamiltonian with the $SU(3)$ symmetry, physical state is described by eight dynamic quantities. They are spin and quadrupole matrix. Two Casimir invariants reduce the number of independent magnetic degrees of freedom down to six. The unitary $SU(3)$ symmetry leads to a modification of the functional hypothesis and includes the quadrupole matrix in the set of thermodynamic quantities. This degree of freedom leads to the appearance of quadrupole spectra of magnetic excitations. In the exchange approximation, the dispersion of spin and quadrupole waves is quadratic function of wave vector \cite{18,19}. In papers \cite{19,20,21}, it was considered the influence of dissipative processes and established the form of relaxation flows, and found damping coefficients of collective excitation spectra.

The concept of a spontaneously broken symmetry \cite{35,36} is one of the most important notion effectively used in the description of equilibrium states of condensed matter. Its application has led to the establishment of strict inequality characterizing the decay of correlations in degenerate states, showed the connection between the broken symmetry and gapless mode in small wave vectors (Goldstone theorem), clarified the relationship of phase transitions and dimensions physical system, and connected the residual symmetry and problem of classification of equilibrium states \cite{36,37,38,39,40,41,42}.

An effective tool for studying magnetic systems is two-time Green's functions \cite{43,44,45}, the knowledge of which allows one to understand both the state of equilibrium and peculiarities of non-equilibrium processes if the deviation from equilibrium is small. As a rule, finding them for specific physical systems uses different kinds of approximate methods. These include, in particular, the quasiparticle approximation, random phase method, and Tyablikov approximation \cite{45,46,47,48,49,50,51}. The above methods have been used in \cite{52,53} for study of high-spin magnets ($s>1/2$).

An important component of study of condensed matter for understanding the experimental data of the internal structure of the matter is the knowledge of two-time Green's functions in the low-frequency range, which is closely related with the behavior of the physical system at large times (hydrodynamic stage of  evolution). Previously, this approach allowed to find Green's functions for a number of superfluid and magnetic condensed media \cite{54,55,56,57,58,60,61,62,63}.

This paper investigates ferromagnets and quadrupole magnets with the exchange $SO(3)$ and $SU(3)$ symmetry of the Hamiltonian. We have calculated low-frequency asymptotics of Green's functions and conducted a comparative analysis for these states. The results can be useful for the study of inelastic scattering of cold neutrons or other particles with the spin $s>1/2$ and determine macroscopic properties of magnetic state.

The paper's structure is as follows: in section \ref{sec2} we describe the Hamiltonian approach for spin $s=1$ magnets. Non-linear dynamic equations with the presence of an external variable field for these systems have been obtained. Section \ref{sec3} describes phase states of magnets with the $SU(3)$ symmetry of the exchange interaction and contains calculations of low-frequency asymptotics of Green's functions of arbitrary local physical quantities for ferro- and quadrupole magnetic states. We have compared them with Green's functions of the ferromagnet with the $SO(3)$ symmetry of the exchange Hamiltonian. Finally, in section \ref{sec5} we discuss the role of unitary symmetry of exchange interaction in structure of low-limit asymptotics of Green's functions.
\section{Magnetic degrees of freedom and mechanics of spin s=1 magnets}\label{sec2}
Description of nonequilibrium processes in magnets in the framework of the Hamiltonian approach involves establishing of a set of dynamic variables that characterize macroscopic state of the system, finding Poisson brackets for them and taking into account symmetry properties of the Hamiltonian. Let us introduce the magnetic degrees of freedom for spin $s=1$ systems according article \cite{18}. To this goal we define the density of the kinematic part of the Lagrangian by the formula: $\textrm{L}(\textbf{x})=b_{\alpha\beta}(\textbf{x})\dot{a}_{\alpha\beta}(\textbf{x})\equiv{tr\hat{b}(\textbf{x})\hat{\dot{a}}(\textbf{x})}$, where $a$  and $b$ are Hermitian $3\times3$ matrices $(\hat{a}=\hat{a}^+,\hat{b}=\hat{a}^+)$. These matrices are canonically conjugate values for which valid Poisson brackets are as follows:
\begin{multline}\label{eq21}
\{b_{\alpha\beta}(\textbf{x}),b_{\mu\nu}(\textbf{x}')\}=0, \{a_{\alpha\beta}(\textbf{x}),a_{\mu\nu}(\textbf{x}')\}=0, \\ \{a_{\alpha\beta}(\textbf{x}),b_{\mu\nu}(\textbf{x}')\}=-\delta_{\alpha\nu}\delta_{\beta\mu}\delta(\textbf{x}-\textbf{x}').
\end{multline}
Here, $\delta_{\alpha\beta}$ is the Kronecker symbol and $\delta(\textbf{x})$ is the Dirac delta function. We connect these matrices with physical quantities of spin $s=1$ magnets. The density of the generator of the $SU(3)$ symmetry is given by relation:
\begin{equation}\label{eq22}
\hat{g}(\textbf{x})\equiv{i[\hat{b}(\textbf{x}),\hat{a}(\textbf{x})]}.
\end{equation}
Square brackets here and below denote the commutator of two matrices. Using (\ref{eq22}) and (\ref{eq21}), we find Poisson bracket for this value
\begin{multline}\label{eq23}
i\{g_{\alpha\beta}(\textbf{x}),g_{\gamma\rho}(\textbf{x}')\}= \\ =(g_{\gamma\beta}(\textbf{x})\delta_{\alpha\rho}-g_{\alpha\rho}(\textbf{x})\delta_{\gamma\beta})\delta(\textbf{x}-\textbf{x}').
\end{multline} 		
As the matrix $\hat{g}(\textbf{x})$ being traceless, two Casimir invariants of the algebra (\ref{eq23}) defined by the expressions: $g_n(\textbf{x})\equiv tr\hat{g}^n(\textbf{x})$, $\{g_n(\textbf{x}),g_{\alpha\beta}(\textbf{x}')\}=0$, where $n=2,3$. The presence of these Casimir invariants reduces the number of independent magnetic degrees of freedom characterizing the macroscopic state of spin $s=1$ magnets down to six.

Under normal multi-sublattice states of magnets and degenerate single-sublattice states, the exchange Hamiltonian is a functional of the matrix $\hat{g}(\textbf{x})$: $H(\hat{g})=\int{d^3xe(\textbf{x},\hat{g}(\textbf{x}'))}$. The exchange energy density is a function of this matrix and its gradient $e(\textbf(x))=e(\hat{g}(\textbf{x}),\nabla\hat{g}(\textbf{x}))$. The stationary action principle leads to Hamiltonian dynamic equations: $\dot{\hat{g}}(\textbf{x})=\{\hat{g}(\textbf{x}),H(\hat{g})\}$. Using (\ref{eq23}) we obtain a functional dynamic equation for the matrix $\hat{g}(\textbf{x})$
\begin{equation}\label{eq24}
\dot{\hat{g}}=i\left[\hat{g}(\textbf{x}),\frac{\delta\hat{H}(\hat{g})}{\delta g(\textbf{x})}\right],
\end{equation}
which generalizes the Landau-Lifshitz equation for the case of magnets with spin $s=1$. In the case of arbitrary spin, canonically conjugate quantities and generator of the $SU(2s+1)$ symmetry can be introduced similar to expressions (\ref{eq21}) and (\ref{eq22}) and the dimension of these matrices is equal to $2s+1$.

Real magnetic degrees of freedom for spin $s=1$ magnets - spin density $s_{\alpha}(\textbf{x})$ and quadrupole matrix $q_{\alpha\beta}(\textbf{x})$ - are connected with the matrix $\hat{g}(\textbf{x})$ by the relation
\begin{equation}\label{eq25}
g_{\alpha\beta}(\textbf{x})\equiv q_{\alpha\beta}(\textbf{x})-i\varepsilon_{\alpha\beta\gamma}s_{\gamma}(\textbf{x})/2.
\end{equation}
The feedback has the form: $s_{\alpha}=i\varepsilon_{\alpha\beta\gamma}g_{\beta\gamma}$, $q_{\alpha\beta}=(g_{\alpha\beta}+g_{\beta\alpha})/2$. The quadrupole matrix is symmetric and traceless tensor: $q_{\alpha\beta}=q_{\beta\alpha}$, $q_{\alpha\alpha}=0$. Five of its independent components can be parameterized as:
\begin{center}
$q_{\alpha\beta}=q_1\left(e_{\alpha}e_{\beta}-\delta_{\alpha\beta}/3\right)+q_2\left(f_{\alpha}f_{\beta}-\delta_{\alpha\beta}/3\right).$
\end{center}
Here, $q_1$, $q_2$ are scalar parameters of this matrix. Vectors $e_{\alpha}, f_{\alpha}, d_{\alpha}=(\textbf{e}\times\textbf{f})_{\alpha}$ form an orthonormal frame and have the physical meaning of magnetic anisotropy axes of the quadrupole ordering. Magnetic states, in which $q_1\neq0, q_2=0$ or $q_2\neq0,q_1=0$, are uniaxial. The case of $q_1\neq0, q_2\neq0$ corresponds to biaxial quadrupole magnetic ordering.

It can be seen that for the vector  $s_{\alpha}(\textbf{x})$, taking into account (\ref{eq23}) and (\ref{eq25}), Poisson bracket
\begin{equation}\label{eq26}
\left\{s_{\alpha}(\textbf{x}),s_{\beta}(\textbf{x}')\right\}=\delta(\textbf{x}-\textbf{x}')\varepsilon_{\alpha\beta\gamma}s_{\gamma}(\textbf{x})
\end{equation}
is true. For the quadrupole matrix due to (\ref{eq23}) and (\ref{eq25}) we get:
\begin{multline}\label{eq27}
\{s_{\alpha}(\textbf{x}),q_{\beta\gamma}(\textbf{x}')\}=\delta(\textbf{x}-\textbf{x}')(\varepsilon_{\alpha\beta\rho}q_{\rho\gamma}(\textbf{x})+\varepsilon_{\alpha\gamma\rho}q_{\rho\beta}(\textbf{x})), \\ \{q_{\alpha\beta}(\textbf{x}),q_{\mu\nu}(\textbf{x}')\}=\delta(\textbf{x}-\textbf{x}')s_{\gamma}(\textbf{x})(\varepsilon_{\gamma\alpha\nu}\delta_{\beta\mu}+ \\ +\varepsilon_{\gamma\beta\mu}\delta_{\alpha\nu}+\varepsilon_{\gamma\beta\nu}\delta_{\alpha\mu}+\varepsilon_{\gamma\alpha\mu}\delta_{\beta\nu})/4.
\end{multline}
The algebra of Poisson brackets (\ref{eq26}), (\ref{eq27}) allows one to describe the dynamics of nonequilibrium states of magnets with the $SU(3)$ symmetry of the Hamiltonian in terms of the spin density and quadrupole matrix. For the set of magnetic degrees of freedom (\ref{eq26}), (\ref{eq27}) it is also possible ferro-quadrupol states, for which both $s_{\alpha}\neq0$ and $q_{\alpha\beta}\neq0$. The subalgebra of Poisson brackets (\ref{eq26}) contains a spin vector. This particular case is analogous to the spin $s=1/2$ systems and shows the validity of the description of the dynamics of magnets with spin $s=1$ by the Landau-Lifshitz equation \cite{13}, if the exchange interaction has the $SO(3)$ symmetry.

In terms of matrices $q_{\alpha\beta}$ and $\varepsilon_{\alpha\beta}\equiv\varepsilon_{\alpha\beta\gamma}s_{\gamma}/2$, the equation (\ref{eq24}) can be represented as two real matrix equations
\begin{multline}\label{eq28}
\dot{\hat{q}}=\left[\hat{\varepsilon},\frac{\delta\hat{H}(\hat{q},\hat{\varepsilon})}{\delta q}\right]-\left[\hat{q},\frac{\delta\hat{H}(\hat{q},\hat{\varepsilon})}{\delta\varepsilon}\right], \\ \dot{\hat{\varepsilon}}=\left[\frac{\delta\hat{H}(\hat{q},\hat{\varepsilon})}{\delta q},\hat{q}\right]+\left[\frac{\delta\hat{H}(\hat{q},\hat{\varepsilon})}{\delta\varepsilon},\hat{\varepsilon}\right].
\end{multline}
We take into account the property of the $SU(3)$ symmetry of the exchange energy density
\begin{center}
$\{\hat{G},e\}=\left[\hat{g},\frac{\partial\hat{e}(\hat{g})}{\partial g}\right]+\left[\nabla_k\hat{g},\frac{\partial\hat{e}(\hat{g})}{\partial\nabla_k g}\right]=0,$
\end{center}
where $\hat{G}\equiv\int{d^3x\hat{g}(\textbf{x})}$ is a generator of $SU(3)$ symmetry. This formula allows to convert equations (\ref{eq28}) to the form of differential conservation laws:
\begin{multline}\label{eq29}
\dot{\hat{q}}=-\nabla_k\left(\left[\hat{\varepsilon},\frac{\delta\hat{e}(\hat{q},\hat{\varepsilon})}{\partial\nabla_k q}\right]+\left[\frac{\delta\hat{e}(\hat{q},\hat{\varepsilon})}{\partial\nabla_k\varepsilon},\hat{q}\right]\right), \\ \dot{\hat{\varepsilon}}=-\nabla_k\left(\left[\frac{\delta\hat{e}(\hat{q},\hat{\varepsilon})}{\partial\nabla_k q},\hat{q}\right]+\left[\frac{\delta\hat{e}(\hat{q},\hat{\varepsilon})}{\partial\nabla_k\varepsilon},\hat{\varepsilon}\right]\right).
\end{multline}

The two-time retarded Green's function for arbitrary quasi-local operators $\hat{a}$ and $\hat{b}$ shall be defined by the relation \cite{43}
\begin{equation}\label{eq210}
G_{ab}(\textbf{x},t;\textbf{x}',t')\equiv-i\theta(t-t')Sp\hat{w}\left[\hat{a}(\textbf{x},t),\hat{b}(\textbf{x}',t')\right].
\end{equation}
Here, $\hat{\omega}$ is the Gibbs equilibrium statistical operator and operators $\hat{a}$, $\hat{b}$ are functionals of Bose creation and annihilation operators in Heisenberg representation. The linear response of the value $a$ to the external disturbance is the following:
\begin{center}
$\delta a_{\xi}(\textbf{x},t)=\int_{-\infty}^{+\infty}{dt'\int{d^3x'\delta\xi(\textbf{x}',t')G_{ab}(\textbf{x}-\textbf{x}',t-t')}}$.
\end{center}
The Fourier transformation of this relation can be written as:
\begin{equation}\label{eq211}
\delta a_{\xi}(\textbf{k},\omega)=G_{ab}(\textbf{k},\omega)\delta\xi(\textbf{k},\omega).
\end{equation}
On the other hand, Hamiltonian formalism allows to obtain equations of macroscopic dynamics of magnetic systems in an external field. The linearized version of these equations connects the deviation of the local physical quantity $\delta a$ and potential of the field $\delta\xi$ thus allowing us to find asymptotics of two-time Green's functions at low frequencies $\omega\tau_r<<1$ and small wave vectors $kl<<1$. We consider the effect of a weak alternating field on the evolution of the studied magnetic system. The complete Hamiltonian of the magnetic system has the form: $H(t)=H+V(t)$. Here, $V(t)$ is the energy of interaction of the magnetic system with an external field:
\begin{equation}\label{eq212}
V(t)=\int{d^3x\delta\xi(\textbf{x},t)b(\textbf{x},t)}.
\end{equation}
Here, $\delta\xi(\textbf{x},t)$ is a potential of interaction of the magnets and the external field, $b(\textbf{x},t)$ is a local physical quantity. We suppose that the external field changing slowly so the characteristic frequency of its changes is small compared to $\tau_r^{-1}$. Here, $\tau_r$ is a time of relaxation (time of establishment of local equilibrium of the magnetic system). In this case, the physical system has time to adapt to instantaneous values of the field. At time $t>>\tau_r$ value $b(\textbf{x},t)$ depends on time through magnetic degrees of freedom, i.e. $b(\textbf{x},t)\underset{t>>\tau_r}{\longrightarrow}b(\textbf{x},\hat{g}(\textbf{x},t))$. For the validity of this formula, spatial scales of changes $l$ of macroscopic values must be greater than the average interatomic distance $l>>a$. In accordance with the above and taking into account (\ref{eq210}), we obtain the dynamic equation for matrix $\hat{g}(\textbf{x})$ in the presence of an external field
\begin{equation}\label{eq213}
\dot{\hat{g}}=i\left[\hat{g},\frac{\delta\hat{H}(\hat{g})}{\delta g}\right]+\hat{\eta}(\hat{g}), \hat{\eta}(\hat{g})=i\delta\xi\left[\hat{g},\frac{\partial\hat{b}(\hat{g})}{\delta g}\right]
\end{equation}
where the right-hand side of the equation contains the source associated with this field. In terms of real quantities – quadrupole matrix and spin density nonlinear equations of the dynamics of spin $s=1$ magnets have the form
\begin{multline}\label{eq214}
\dot{s}_{\alpha}=-\varepsilon_{\alpha\beta\gamma}\nabla_k\left(\left[\frac{\delta\hat{e}}{\partial\nabla_k q},\hat{q}\right]+\left[\frac{\delta\hat{e}}{\partial\nabla_k\varepsilon},\hat{\varepsilon}\right]\right)_{\beta\gamma}+\eta_{\alpha}(\textbf{s}), \\
\dot{\hat{q}}=-\nabla_k\left(\left[\hat{\varepsilon},\frac{\delta\hat{e}}{\partial\nabla_k q}\right]+\left[\frac{\delta\hat{e}}{\partial\nabla_k\varepsilon},\hat{q}\right]\right)+\hat{\eta}(\hat{q}).
\end{multline}
For sources in equations (\ref{eq214}), the following expressions are obtained:
\begin{center}
$\eta_{\alpha}(\textbf{s})=\delta\xi\varepsilon_{\alpha\beta\gamma}\left(\frac{\partial b}{\partial s_{\beta}}s_{\gamma}+2\frac{\partial b}{\partial q_{\beta\sigma}}q_{\gamma\sigma}\right)$,

$\eta_{\beta\gamma}(\hat{q})=-\delta\xi\frac{\partial b}{\partial s_{\alpha}}(q_{\beta\rho}\varepsilon_{\alpha\gamma\rho}+q_{\rho\gamma}\varepsilon_{\alpha\beta\rho})+\delta\xi\frac{\partial b}{\partial q_{\mu\nu}}s_{\sigma}(\varepsilon_{\sigma\beta\nu}\delta_{\gamma\mu}+\varepsilon_{\sigma\gamma\nu}\delta_{\beta\mu})/2$.
\end{center}

Despite the small  spatial inhomogeneities, the magnetic system can be both as close to and as far from equilibrium states. In the following sections, where low-frequency asymptotics of Green's functions are calculated, we suppose that deviations from equilibrium are small.
\section{Dynamics of quadrupole and ferromagnetic states in an external field. Low-frequency asymptotics of Green's functions}\label{sec3}
The analytical form of the $SU(3)$ symmetrical exchange Hamiltonian can be constructed by analogy with the Heisenberg Hamiltonian. We consider the exchange energy in the form \cite{18}:
\begin{equation}\label{eq31}
H=-\int{d^3xd^3x'J(|\textbf{x}-\textbf{x}'|)tr\hat{g}(\textbf{x})\hat{g}(\textbf{x}')}.
\end{equation}
Here, $J(|\textbf{x}-\textbf{x}'|)$ is an exchange integral of two-particle magnetic interaction. The quadratic approximation of spatial inhomogeneities leads to the energy density expression $e(\textbf{x})=-Jg_2(\textbf{x})+\bar{J}tr\nabla\hat{g}(\textbf{x})\nabla\hat{g}(\textbf{x})/2$. Constants of homogeneous exchange $J$ and inhomogeneous exchange $\bar{J}$ in the energy density are connected with the exchange integral by the following relations: $J=\int{d^3xJ(|\textbf{x}|)}$ and $\bar{J}=\int{d^3xx^2J(|\textbf{x}|)/3}$. The functional form of the term of the homogeneous exchange is defined by the Casimir invariant $g_2=s^2/2+2q^2/3>0$. We present the model of the exchange energy as a sum of two terms:
\begin{multline}\label{eq32}
e=e_0+e_n, e_0(s,q)=-Jg_2+Bg_2^2+Aq^2, \\ e_n=\bar{J}tr(\nabla_k\hat{g})^2/2.
\end{multline}
The first two terms in the homogeneous exchange energy possess the $SU(3)$ symmetry and the last term possesses the $SO(3)$ symmetry. It is necessary for the existence of nontrivial solutions of magnetic degrees of freedom in equilibrium. For definiteness, we assume that the quadrupole matrix is uniaxial. The density of the inhomogeneous exchange energy possesses the $SU(3)$ symmetry and is positive for case $\bar{J}>0$.

Equilibrium values of spin modules and quadrupole matrix as well as the stability of magnetic states we obtain from the conditions
\begin{multline}\label{eq33}
\partial e_0/\partial s=0, \partial e_0/\partial q=0, \partial^2e_0/\partial s^2>0, \\ \partial^2e_0/\partial q^2>0, \frac{\partial^2e_0}{\partial s^2}\frac{\partial^2e_0}{\partial q^2}-\left(\frac{\partial^2e_0}{\partial s\partial q}\right)^2>0.
\end{multline}
The system of equations (\ref{eq33}) has three solutions: 1. paramagnet: $s_0=q_0=0$ is stable if $J<0$ and $3A>2J$. 2. quadrupole magnet: $s_0=0$ and $q_0^2=3(2J-3A)/8B>0$. The state exists and is stable if $A<0$, $B>0$, $2J>3A$. 3. ferromagnet: $s_0^2=J/B>0$ and $q_0=0$. This state is stable if $A>0$, $B>0$, $J>0$. For the exchange energy (\ref{eq32}), equations (\ref{eq214}) can be simplified and take the form:
\begin{multline}\label{eq34}
\dot{s}_{\alpha}=\varepsilon_{\alpha\beta\gamma}\bar{J}\left(\left[\hat{q},\triangle\hat{q}\right]+\left[\triangle\hat{\varepsilon},\hat{\varepsilon}\right]\right)_{\beta\gamma}+\hat{\eta}(\textbf{s})_{\alpha}, \\ \dot{\hat{q}}=\bar{J}\left[\triangle\hat{\varepsilon},\hat{q}\right]+\bar{J}\left[\triangle\hat{q},\hat{\varepsilon}\right]+\hat{\eta}(\hat{q}).
\end{multline}
These equations describe the dynamics of the studied magnetic system in the presence of external field. The linearization of these equations does not allow us to obtain expression of the asymptotics of Green's functions corresponding to the paramagnetic state and requires taking into account relaxation processes.

We calculate the asymptotics of Green's functions for the equilibrium state corresponding to the case 2. For the quadrupole magnetic state, basing on (\ref{eq34}), we obtained linearized equations:
\begin{center}
$\delta\dot{\hat{q}}=\bar{J}[\triangle\delta\hat{\varepsilon},\hat{q}_0]+\hat{\eta}(\hat{q}),$
\end{center}
\begin{center}
$\eta_{\beta\gamma}(\hat{q})=-\delta\xi\frac{\partial b}{\partial s_{\alpha}}(\varepsilon_{\alpha\beta\rho}q^0_{\rho\gamma}+\varepsilon_{\alpha\gamma\rho}q^0_{\rho\beta}),$
\end{center}
\begin{center}
$\delta\dot{s}_{\alpha}=\bar{J}\varepsilon_{\alpha\beta\gamma}[\hat{q}_0,\triangle\delta\hat{q}]_{\beta\gamma}+\eta_{\alpha}(\textbf{s}),$
\end{center}
\begin{center}
$\eta_{\alpha}(\textbf{s})=2\delta\xi\varepsilon_{\alpha\beta\gamma}\frac{\partial b}{\partial q_{\beta\lambda}}q^0_{\gamma\lambda}.$
\end{center}
In terms of the Fourier representation, we find:
\begin{multline}\label{eq35}
i\omega\delta\hat{q}=-k^2\bar{J}[\delta\hat{\varepsilon},\hat{q}_0]+\hat{\eta}(\hat{q}), \\ i\omega\delta s_{\alpha}=-k^2\bar{J}\varepsilon_{\alpha\beta\gamma}[\hat{q}_0,\delta\hat{q}]_{\beta\gamma}+\eta_{\alpha}(\textbf{s}).
\end{multline}
Excluding the variation of the quadrupole matrix using the second equation in (\ref{eq35}), we connected the variation of the spin density $\delta s_{\lambda}=D^{-1}_{\lambda\alpha}\bar{\eta}_{\alpha}$ with external field potential. The matrix $D_{\alpha\beta}$ and the source $\bar{\eta}_{\alpha}$ are equal
\begin{multline}\label{eq36}
D_{\alpha\beta}=i\omega\delta_{\alpha\beta}+ik^4\bar{J}^2[3\hat{q}^2_0-2\hat{I}tr\hat{q}^2_0]_{\alpha\beta}/\omega, \\ \bar{\eta}_{\alpha}\equiv\eta_{\alpha}(\textbf{s})+2ik^2\delta\xi\bar{J}\frac{\partial b}{\partial s_{\beta}}[3\hat{q}^2_0-2\hat{I}tr\hat{q}^2_0]_{\alpha\beta}/\omega.
\end{multline}
We assume that the quadrupole matrix is uniaxial in the equilibrium state $q^0_{\alpha\beta}=q_0(l_{\alpha}l_{\beta}-\delta_{\alpha\beta}/3)$, where $l_{\alpha}$ is the axis of the magnetic anisotropy. The response of the value $\delta a(\textbf{k},\omega)$ on external field $\delta\xi(\textbf{k},\omega)$ in the main approximation of small wave vectors and frequencies have the form:
\begin{center}
$\delta a(\textbf{k},\omega)=\frac{\partial a}{\partial s_{\alpha}}\delta s_{\alpha}(\textbf{k},\omega)+\frac{\partial a}{\partial q_{\alpha\beta}}\delta q_{\alpha\beta}(\textbf{k},\omega)$.
\end{center}
Then, using formulas (\ref{eq36}) and (\ref{eq214}), we obtain the expression of asymptotics of two-time Green's function in terms of basic Green's functions:
\begin{center}
$G_{ab}(\textbf{k},\omega)=\frac{\partial a}{\partial s_{\alpha}}G_{s_{\alpha},s_{\beta}}(\textbf{k},\omega)\frac{\partial b}{\partial s_{\beta}}+\frac{\partial a}{\partial q_{\alpha\beta}}G_{q_{\alpha\beta},q_{\gamma\rho}}(\textbf{k},\omega)\frac{\partial b}{\partial q_{\gamma\rho}}+\frac{\partial a}{\partial s_{\alpha}}G_{s_{\alpha},q_{\mu\lambda}}(\textbf{k},\omega)\frac{\partial b}{\partial q_{\mu\lambda}}+\frac{\partial a}{\partial q_{\mu\lambda}}G_{q_{\mu\lambda},s_{\alpha}}(\textbf{k},\omega)\frac{\partial b}{\partial s_{\alpha}}$.
\end{center}
For these functions, taking into account (\ref{eq36}), we find expressions:
\begin{multline}\label{eq37}
G_{s_{\alpha},s_{\beta}}(\textbf{k},\omega)=-\frac{2\bar{J}k^2q^2_0\delta_{\alpha\beta}^{\bot}(\textbf{l})}{\triangle(\textbf{k},\omega)}, \\ G_{q_{\alpha\beta},q_{\gamma\rho}}(\textbf{k},\omega)=\frac{\bar{J}k^2q^2_0F_{\alpha\beta,\gamma\rho}(\textbf{l})}{2\triangle(\textbf{k},\omega)},\\ G_{q_{\beta\gamma},s_{\alpha}}(\textbf{k},\omega)=\frac{i\omega q_0F_{\beta\gamma}^{\alpha}(\textbf{l})}{\triangle(\textbf{k},\omega)}=-G_{s_{\alpha},q_{\beta\gamma}}(\textbf{k},\omega).
\end{multline}
Here notations are introduced:
\begin{center}
$\triangle(\textbf{k},\omega)=\omega^2-k^4\bar{J}^2q_0^2$, \\ $F_{\alpha\beta,\gamma\rho}(\textbf{l})=\delta_{\alpha\gamma}^{\bot}(\textbf{l})l_{\beta}l_{\rho}+\delta_{\alpha\rho}^{\bot}(\textbf{l})l_{\beta}l_{\gamma}+\delta_{\gamma\beta}^{\bot}(\textbf{l})l_{\alpha}l_{\rho}+\delta_{\rho\beta}^{\bot}(\textbf{l})l_{\alpha}l_{\gamma}$, \\ $F_{\beta\gamma}^{\alpha}(\textbf{l})\equiv\varepsilon_{\alpha\beta}(\textbf{l})l_{\gamma}+\varepsilon_{\alpha\gamma}(\textbf{l})l_{\beta}$, \\ $\delta_{\mu\lambda}^{\bot}(\textbf{l})\equiv\delta_{\mu\lambda}-l_{\mu}l_{\lambda}, \varepsilon_{\mu\lambda}(\textbf{l})=\varepsilon_{\mu\lambda\rho}l_{\rho}$.
\end{center}

Let's consider some particular cases of asymptotics (\ref{eq37}). At $\omega=0, k\neq0$, we can see that, in accordance with the Bogolyubov's theorem \cite{35}, Green's functions $G_{s_{\alpha},s_{\beta}}(\textbf{k},0)$ and $G_{q_{\alpha\beta},q_{\gamma\rho}}(\textbf{k},0)$ have a singularity $1/k^2$: $G_{s_{\alpha},s_{\beta}}(\textbf{k},0)=\frac{2\delta_{\alpha\beta}^{\bot}(\textbf{l})}{\bar{J}k^2}$, $G_{q_{\alpha\beta},q_{\gamma\rho}}(\textbf{k},0)=\frac{F_{\alpha\beta,\gamma\rho}(\textbf{1})}{4\bar{J}k^2}$ and the Green's function $G_{s_{\alpha},q_{\beta\gamma}}(\textbf{k},0)=0$ vanishes. If the wave vector $\textbf{k}=0$ and $\omega\neq0$, basic Green's functions $G_{s_{\alpha},s_{\beta}}(0,\omega)=G_{q_{\alpha\beta},q_{\gamma\rho}}(0,\omega)=0$ vanish and the Green's function $G_{s_{\alpha},q_{\beta\gamma}}(0,\omega)$ has a singularity in frequency $G_{s_{\alpha},q_{\beta\gamma}}(0,\omega)=-iF_{\beta\gamma}^{\alpha}(\textbf{l})/\omega$. From condition $\triangle(\textbf{k},\omega)=0$ we get a quadrupole wave spectrum $\omega=k^2\bar{J}q_0$ propagating near the equilibrium state corresponding to uniaxial quadrupole ordering.

We shall now calculate asymptotics of Green's functions for magnets corresponding to the case 3. Close to the ferromagnetic state ($s_0\neq0,\hat{q}_0=0$), linearized dynamic equations (\ref{eq214}) have the form
\begin{multline}\label{eq38}
\delta\dot{s}_{\alpha}=\varepsilon_{\alpha\beta\gamma}\bar{J}[\triangle\delta\hat{\varepsilon},\hat{\varepsilon}_0]_{\beta\gamma}+\eta_{\alpha}(\textbf{s}), \\ \eta_{\alpha}(\textbf{s})=\delta\xi\varepsilon_{\alpha\beta\gamma}\frac{\partial b}{\partial s_{\beta}}s^0_{\gamma}, \\ \delta\dot{\hat{q}}=\bar{J}[\triangle\delta\hat{q},\hat{\varepsilon}_0]+\hat{\eta}(\hat{q}), \\ \eta_{\beta\gamma}(\hat{q})=\delta\xi\frac{\partial b}{\partial q_{\mu\nu}}s^0_{\sigma}(\varepsilon_{\sigma\beta\nu}\delta_{\gamma\mu}+\varepsilon_{\sigma\gamma\nu}\delta_{\beta\mu}+(\mu\leftrightarrow\nu))/4.
\end{multline}
We see that dynamic equations for the spin density and quadrupole matrix in this case are separated. In terms of the Fourier representation, the first equation (\ref{eq38}) leads to the relation of the variation of spin density with the external field potential
\begin{equation}\label{eq39}
\delta s_{\alpha}(\textbf{k},\omega)=D_{\alpha\beta}^{-1}(\textbf{k},\omega)\eta_{\beta}(\textbf{s};\textbf{k},\omega),
\end{equation}
where
\begin{multline}\label{eq310}
D_{\alpha\beta}(\textbf{k},\omega)=i\omega\delta_{\alpha\beta}-\bar{J}k^2\varepsilon_{\alpha\beta\gamma}s_{\gamma}^0/2, \\ \eta_{\beta}(\textbf{s};\textbf{k},\omega)=\delta\xi(\textbf{k},\omega)\varepsilon_{\beta\mu\nu}\frac{\partial b}{\partial s_{\mu}}s^0_{\nu}.
\end{multline}
The second equation (\ref{eq38}) leads to the linearized equation for the variation of the quadrupole matrix
\begin{multline}\label{eq311}
i\omega\delta q_{\beta\gamma}(\textbf{k},\omega)=\bar{J}k^2s_0(\varepsilon_{\beta\rho3}\delta q_{\rho\gamma}-\varepsilon_{\rho\gamma3}\delta q_{\rho\beta})/2+\eta_{\beta\gamma}(\hat{q};\textbf{k},\omega), \\ \eta_{\beta\gamma}(\textbf{k},\omega)=\delta\xi(\hat{q};\textbf{k},\omega)s_0\left(\frac{\partial b}{\partial q_{\gamma\nu}}+\frac{\partial b}{\partial q_{\nu\gamma}}\right)\varepsilon_{3\beta\nu}/4+ \\ \delta\xi(\hat{q};\textbf{k},\omega)s_0\left(\frac{\partial b}{\partial q_{\beta\nu}}+\frac{\partial b}{\partial q_{\nu\beta}}\right)\varepsilon_{3\gamma\nu}/4.
\end{multline}
Here, $s_{\alpha}^0=s_0(0,0,1)$. The equation (\ref{eq311}) contains a double summation of spin indices. To solve it, it is convenient to switch to new variables and single summation by substituting: $q_{\alpha\beta}\rightarrow q_n$,$\eta_{\alpha\beta}\rightarrow\eta_n$, $n=1,2,..,9$: $q_{11}\rightarrow q_1, q_{12}\rightarrow q_2,.., q_{33}\rightarrow q_9$. Such substitution leads to the separation of equations for the quantities $\delta q_{m}$ with indices $(m=1,2,4,5)$, quantities $\delta q_p$ with indices $(p=3,6,7,8)$, and $\delta q_9$. The result is the expression of variations in terms of the external field potential:
\begin{multline}\label{eq312}
\delta q_m(\textbf{k},\omega)=\delta\xi(\textbf{k},\omega)s_0D_{mn}^{-1}(\textbf{k},\omega)M_{nl}\frac{\partial b}{\partial q_l}/2, \\ m,n,l=1,2,4,5; \\ \delta q_p(\textbf{k},\omega)=\delta\xi(\textbf{k},\omega)s_0\underline{D}_{ps}^{-1}(\textbf{k},\omega)\underline{M}_{st}\frac{\partial b}{\partial q_t}/4, \\ p,s,t=3,6,7,8; \\ \delta q_9(\textbf{k},\omega)=0.
\end{multline}
Matrices $\hat{D}, \hat{M}, \underline{\hat{D}}, \underline{\hat{M}}$ in the right side of expressions (\ref{eq312}) are presented in the form of a direct product of Pauli matrices $\hat{\sigma}_{\alpha}$ and unit matrix $\hat{I}_2$:
\begin{center}
$\hat{D}(\textbf{k},\omega)=i\omega\hat{I}_4+i\kappa(\textbf{k})\hat{M}$, \\ $\hat{M}=\hat{\sigma}_2\otimes\hat{I}_2+\hat{I}_2\otimes\hat{\sigma}_2,$ \\ $\hat{\underline{D}}(\textbf{k},\omega)=i\omega\hat{I}_4-\kappa(\textbf{k})\hat{I}_2\otimes\hat{\sigma}_2,$ \\ $\hat{\underline{M}}=\hat{\sigma}_1\otimes\hat{\sigma}_2+\hat{I}_2\otimes\hat{\sigma}_2$.
\end{center}
Here the notation $\kappa(\textbf{k})\equiv k^2s_0\bar{J}/2$ is introduced. We present the variation of the local physical value near the ferromagnetic state as
\begin{center}
$\delta a(\textbf{k},\omega)=\frac{\partial a}{\partial s_{\alpha}}\delta s_{\alpha}(\textbf{k},\omega)+\frac{\partial a}{\partial q_n}\delta q_n(\textbf{k},\omega)+\frac{\partial a}{\partial q_s}\delta q_s(\textbf{k},\omega)$.
\end{center}
Taking into account formulas (\ref{eq39})-(\ref{eq312}) and the equation (\ref{eq212}), we obtained the structure of low-frequency asymptotics of Green's functions for arbitrary local variables $a$ and $b$ in terms of basic Green's functions:
\begin{center}
$G_{ab}(\textbf{k},\omega)=\frac{\partial a}{\partial s_{\alpha}}G_{s_{\alpha},s_{\beta}}(\textbf{k},\omega)\frac{\partial b}{\partial s_{\beta}}+\frac{\partial a}{\partial q_m}G_{q_m,q_n}(\textbf{k},\omega)\frac{\partial b}{\partial q_n}+\frac{\partial a}{\partial q_p}G_{q_p,q_s}(\textbf{k},\omega)\frac{\partial b}{\partial q_s}$.
\end{center}
The explicit forms of these basic functions are:
\begin{multline}\label{eq313}
G_{s_{\alpha},s_{\beta}}(\textbf{k},\omega)=\frac{-i\omega\varepsilon_{\alpha\beta}(\textbf{m})s_0+\bar{J}k^2s^2_0\delta_{\alpha\beta}^{\perp}(\textbf{m})/2}{\triangle_0(\textbf{k},\omega)}, \\ G_{q_m,q_n}(\textbf{k},\omega)=s_0X_{mn}(\textbf{k},\omega)/2\triangle_1(\textbf{k},\omega), \\ G_{q_p,q_s}(\textbf{k},\omega)=s_0X_{ps}(\textbf{k}/2,\omega)/4\triangle_0(\textbf{k},\omega).
\end{multline}
Here, $\textbf{m}\equiv\textbf{s}_0/s_0$ and $4\times4$ matrix $\hat{X}(\textbf{k}/2,\omega)$ takes the form:
\begin{multline}\label{eq314}
\hat{X}(\textbf{k},\omega)=\omega(\hat{\sigma}_2\otimes\hat{I}_2+\hat{I}_2\otimes\hat{\sigma}_2)- \\ 2\kappa(\textbf{k})(\hat{I}_4+\hat{\sigma}_2\otimes\hat{\sigma}_2).
\end{multline}
Denominators in (\ref{eq313}) are determined by formulas $\triangle_0(\textbf{k},\omega)\equiv\omega^2-\kappa^2(\textbf{k})$, and  $\triangle_1(\textbf{k},\omega)\equiv\omega^2-4\kappa^2(\textbf{k})$. It is easy to see that singularities of basic Green's functions (\ref{eq313}) can be of two types. Nonvanishing Green's functions $G_{s_{\alpha},s_{\beta}}, G_{q_{(3,6,7,8)},q_s}$ contain a singularity at $\triangle_0(\textbf{k},\omega)=0$, and Green's functions $G_{q_{(1,2,3,5)},q_s}$ – at $\triangle_1(\textbf{k},\omega)=0$. These two equations lead to the spectra of collective magnetic excitations $\omega=\kappa(\textbf{k})$ and $\omega=2\kappa(\textbf{k})$. Formulas (\ref{eq313}), (\ref{eq314}) solve the problem of finding of low-frequency asymptotics of Green's functions for ferromagnetic states with the spin $s=1$ if the exchange interaction possesses a $SU(3)$ symmetry. We present a special cases of basic Green's functions:
\begin{center}
$G_{s_{\alpha},s_{\beta}}(\textbf{k},o)=-\frac{2\delta_{\alpha\beta}^{\bot}(\textbf{m})}{\bar{J}k^2}$, \\ $G_{s_{\alpha},s_{\beta}}(0,\omega)=-\frac{is_0\varepsilon_{\alpha\beta}(\textbf{m})}{\omega}$, \\ $G_{q_{p},q_{s}}(\textbf{k},o)=\frac{(\hat{I}_4+\hat{\sigma}_2\otimes\hat{\sigma}_2)_{ps}}{2k^2\bar{J}}$, \\ $G_{q_{m},q_{n}}(\textbf{k},o)=\frac{(\hat{I}_4+\hat{\sigma}_2\otimes\hat{\sigma}_2)_{mn}}{2k^2\bar{J}}$, \\ $G_{q_{p},q_{s}}(0,\omega)=\frac{s_0}{4\omega}(\hat{I}_2\otimes\hat{\sigma}_2+\hat{\sigma}_2\otimes\hat{I}_2)_{ps},$ \\ $G_{q_{m},q_{n}}(0,\omega)=\frac{s_0}{2\omega}(\hat{I}_2\otimes\hat{\sigma}_2+\hat{\sigma}_2\otimes\hat{I}_2)_{mn}$.
\end{center}
Asymptotics of Green's functions $G_{s_{\alpha},s_{\beta}}(0,\omega)$ and $G_{s_{\alpha},s_{\beta}}(\textbf{k},o)$ coincide with results provided in \cite{64}.

Now we consider a similar problem of finding low-frequency asymptotics of Green's functions for an $SO(3)$ symmetric exchange interaction with spin $s=1/2$ and compare them with an explicit form of Green's functions (\ref{eq313}) of the $SU(3)$ symmetric ferromagnet. The density of the inhomogeneous exchange energy with an $SO(3)$ symmetry has the form $e_n=\bar{J}(\nabla_ks_{\alpha})^2/4$. This expression follows from (\ref{eq32}), where we neglect quadrupole degrees of freedom. Since $G_{ab}(\textbf{k},\omega)=\frac{\partial a}{\partial s_{\alpha}}\frac{\delta s_{\alpha}(\textbf{k},\omega)}{\delta\xi(\textbf{k},\omega)}$, then according to (\ref{eq39}), (\ref{eq310}), we find
\begin{equation}\label{eq315}
G_{ab}(\textbf{k},\omega)=\frac{\partial a}{\partial s_{\alpha}}G_{s_{\alpha},s_{\beta}}(\textbf{k},\omega)\frac{\partial b}{\partial s_{\beta}}.
\end{equation}
The form of the basic Green's function for the spin density component coincides with the formula (\ref{eq313}) in this case. We use the expression of the Green's function (\ref{eq315}) for the ferromagnetic state and based on it calculate Green's functions $G_{q_{\mu\nu},q_{\beta\gamma}}(\textbf{k},\omega), G_{s_{\alpha},q_{\beta\gamma}}(\textbf{k},\omega)$, noting that $q_{\mu\nu}=s_{\mu}s_{\nu}-s^2\delta_{\mu\nu}/3$. It can be seen that these expressions have the following form:
\begin{multline}\label{eq316}
G_{q_{\mu\nu},q_{\beta\gamma}}(\textbf{k},\omega)=\frac{s_0^3(-i\omega \Phi_{\mu\nu,\beta\gamma}(\textbf{m})+\bar{J}k^2s_0F_{\mu\nu,\beta\gamma}(\textbf{m}))}{\triangle_0(\textbf{k},\omega)}, \\ G_{s_{\alpha},q_{\beta\gamma}}(\textbf{k},\omega)= \\ s_0^2\frac{-i\omega F_{\beta\gamma}^{\alpha}(\textbf{m})+\bar{J}k^2s_0(\delta_{\alpha\beta}^{\bot}(\textbf{m})m_{\gamma}+\delta_{\alpha\gamma}^{\bot}(\textbf{m})m_{\beta})}{\triangle_0(\textbf{k},\omega)}.
\end{multline}
Here we use the nonation
\begin{center}
$\Phi_{\mu\nu,\beta\gamma}(\textbf{m})\equiv\varepsilon_{\mu\beta}(\textbf{m})m_{\nu}m_{\gamma}+\varepsilon_{\mu\gamma}(\textbf{m})m_{\nu}m_{\beta}+\varepsilon_{\nu\beta}(\textbf{m})m_{\mu}m_{\gamma}+\varepsilon_{\nu\gamma}(\textbf{m})m_{\mu}m_{\beta}$.
\end{center}
Special cases of Green's functions (\ref{eq316}) are given by expressions
\begin{multline}\label{eq317}
G_{q_{\mu\nu},q_{\beta\gamma}}(\textbf{k},0)=-\frac{4s_0^2F_{\mu\nu,\beta\gamma}(\textbf{m})}{k^2\bar{J}}, \\ G_{q_{\mu\nu},q_{\beta\gamma}}(0,\omega)=-\frac{is_0^3\Phi_{\mu\nu,\beta\gamma}(\textbf{m})}{\omega}, \\ G_{s_{\alpha},q_{\beta\gamma}}(\textbf{k},0)=-\frac{s_0(\delta_{\alpha\beta}^{\bot}(\textbf{m})m_{\gamma}+\delta_{\alpha\gamma}^{\bot}(\textbf{m})m_{\beta})}{k^2\bar{J}}, \\ G_{s_{\alpha},q_{\beta\gamma}}(0,\omega)=-\frac{is_0^2F_{\beta\gamma}^{\alpha}(\textbf{m})}{\omega}.
\end{multline}
Comparison of formulas (\ref{eq313}) and (\ref{eq317}) shows qualitative agreement of Green's functions singularities in wave vector and frequency at different unitary symmetry of exchange interaction. However, the coefficients of these singularities have an essentially different dependence from equilibrium magnetic values. Besides, if the exchange interaction possesses $SU(3)$ symmetry, then for ferromagnetic state the Green's function $G_{s_{\alpha},q_n}(\textbf{k},\omega)$ vanishes, whereas in the case of $SO(3)$ symmetry of the ferromagnetic exchange interaction, this value is represented by formula (\ref{eq316}). Finally, singularity of Green's functions with $SO(3)$ symmetry of the ferromagnet interaction demonstrates the presence of one branch of the spin waves, while Green's functions with $SU(3)$ symmetric interaction have singularities of two types, reflecting the existence of spin and quadrupole waves.
\section{Discussion and conclusions}\label{sec5}
In this paper, we have investigated the problem of the influence of a weak variable field on non-equilibrium processes of magnets with the spin $s=1$ and obtained nonlinear dynamic equations, which account for the unitary group $SU(2)\sim SO(3)$ or $SU(3)$ symmetries of exchange interactions. Based on these equations, we have found hydrodynamic asymptotics of two-time Green's functions in an explicit form in wave vectors and frequencies. The presented results show the influence of a particular unitary symmetry of the exchange interaction on macroscopic properties of studied magnets. A certain unitary symmetry when realized substantially affects possible magnetic equilibrium states and the nature of the anisotropy of asymptotics of Green's functions in wave vectors and frequencies.

The methodological peculiarity of finding asymptotics of Green's functions is their determination for arbitrary local physical quantities. This makes it possible to conduct a comparative analysis of Green's functions found for magnets with a different unitary symmetry of the exchange interaction. The example of magnetic state (quadrupole magnet) demonstrates peculiarities of additional degrees of freedom, which are even with respect to time reversal. We have used Casimir invariants of the unitary symmetry generators algebra for constructing of the homogeneous part of the exchange energy. The presented approach allows us to consider also the degenerate state in multi-sublattice spin $s=1$ magnets. These states contain additional dynamic variables: antiferromagnet vector or spin nematic order parameter.

The performed study can be usefull in understanding of collective properties of magnetic states and their experimental discovery in spin $s=1$ magnets. The above scheme of accounting of $SO(3)$ or $SU(3)$ symmetries of the magnetic exchange interaction can be generalized to an arbitrary unitary group of the $SU(n)$ symmetry, and, in particular, be used to describe collective properties of magnets with the spin $s=3/2$.

\end{document}